\documentclass[prl, twocolumn, showpacs, amsmath, amssymb, superscriptaddress, aps]{revtex4}

\usepackage{graphicx}
\usepackage{dcolumn}
\usepackage{color}
\usepackage{bm}
\def\he4{$^4$He}
\def\Am3{\AA$^{-3}$}

\newcommand{\dR}{d {\pmb R}}
\newcommand{\Rvec}{{\pmb R}}
\newcommand{\rvec}{{\pmb r}}
\def\beq{\begin{equation}}
\def\eeq{\end{equation}}
\begin{document}

\author{P. Corboz}
\affiliation{Theoretische Physik, ETH Zurich, 8093 Zurich, Switzerland}

\author{L. Pollet}
\affiliation{Theoretische Physik, ETH Zurich, 8093 Zurich, Switzerland}

\author{N. V. Prokof'ev}
\affiliation{Department of Physics, University of Massachusetts,
Amherst, MA 01003, USA}
\affiliation{Russian Research Center ``Kurchatov Institute'', 123182 Moscow, Russia}
\affiliation{Theoretische Physik, ETH Zurich, 8093 Zurich, Switzerland}

\author{M. Troyer}
\affiliation{Theoretische Physik, ETH Zurich, 8093 Zurich, Switzerland}

\title{Binding of a $^3$He impurity to a screw dislocation in solid \he4}

\date{\today}

\begin{abstract}
Using first-principle simulations for the probability density of finding a $^3$He atom in the
vicinity of the screw dislocation in solid \he4,
we determine the binding energy to the dislocation nucleus $E_{\rm B} = 0.8 \pm 0.1 $K
and the density of localized states at larger distances. The specific heat due to $^3$He features a peak
similar to the one observed in recent experiments, and our model can also account for the
observed increase in shear modulus at low temperature. We further discuss the role of $^3$He
in the picture of superfluid defects.

\end{abstract}



\maketitle
The observation of a non-classical moment of rotational inertia (NCRI) in torsional oscillator experiments~\cite{KC04} in solid $^4$He has revived the debate on supersolids, but its understanding proved to be challenging. Several groups over the world have confirmed NCRI, but aspects such as the pressure dependence~\cite{KC04, KC06}, disorder~\cite{Rittner06, Balibar_review}, history dependence (hysteresis)~\cite{Aoki, Clark08}, ``critical" velocity~\cite{KC04, Aoki}, crystal growth~\cite{Balibar_review},  oscillation frequency~\cite{Aoki}, rim velocity, and  $^3$He concentration~\cite{Kim08} all show unexpected behavior and defy any simple physical picture (for a review, see \cite{Balibar_review, Prokofev_review}). Theoretically, consensus is growing towards a network of superfluid defects as the mechanism of superflow, but some effects cannot yet be explained properly, especially at the quantitative level~\cite{Rittner08}.

One of the main puzzles is the effect of even minute $^3$He concentrations.
There is mounting evidence that the interplay between $^3$He impurities and crystallographic defects (dislocations in single crystals) is not innocuous and can, in fact, be understood theoretically to a large degree.
In this Letter, we investigate this topic and focus on the $^3$He binding to screw dislocations, which are
common crystal defects in solid $^4$He.

Day and Beamish observed an increase in the shear modulus of the \he4 crystal when the temperature is lowered~\cite{Day07}, which could be understood from binding of $^3$He to dislocations at low temperatures. According to the Granato-L{\"u}cke theory~\cite{Granato56}, dislocations move in response to shear stress in their glide plane. More precisely, they bow out between pinning centers provided by impurities or intersections, which can reduce the shear modulus by $30\%$ from its intrinsic value in a frequency independent way. When $^3$He binds to a dislocation, it acts as an additional pinning center. Since the change in shear modulus is quadratic in the length between the pinning centers, the shear modulus quickly recovers its intrinsic value. Remarkably, the shear modulus dependence on temperature is nearly identical to that of NCRI.
Yet, the two phenomena are distinct: the NCRI signal can not be fully accounted for by the elasticity effect, nor can the reduction in NCRI by a factor of 100 when blocking an annulus be explained by elasticity arguments~\cite{KC04, Rittner08}.

In the torsional oscillator experiments by  Kim {\it et al.} ~\cite{Kim08} a minimum $^3$He concentration of the order of $x_3\sim 1$ppb seems needed in order to observe NCRI. Then, NCRI increases until $x_3 \sim 1$ppm where a maximum is reached and finally disappears again for concentrations of about 100ppm.
Specific heat measurements showed a nearly constant term in the specific heat at low temperatures scaling with the $^3$He concentration~\cite{Lin07}. After subtraction of the phonon contribution and the mysterious constant term, a peak in the specific heat was found around $T = 0.06$K, which was claimed to be independent of $x_3$ and indicative of the supersolid transition~\cite{Lin07} (see however \cite{Lincorrection}).
In this Letter we show, however, that binding of $^3$He impurities to dislocations results in a
specific heat peak in the same temperature range.

Our approach is numerical and based on Feynman's path-integral formulation of quantum mechanics. The integrals over the paths (world lines) are efficiently evaluated by the worm algorithm \cite{worm}, which has been successful in describing properties of crystallographic defects in solid \he4~\cite{Boninsegni06, Pollet07, Boninsegni07, Pollet08}. We now describe how the method needs to be modified in order to deal with $^3$He impurities.

The grand partition function $Z=\text{Tr}\; {e^{-\beta (\hat H - \mu \hat N)}}$ is expressed as a path integral with the usual discretization of the imaginary time (inverse temperature) $\beta$ into $M$ slices ($\delta=\beta/M$),
\begin{equation}
\label{eq:Z}
Z \approx   \sum_{N=0}^{\infty} e^{\beta \mu N} \int \dR \: T(\Rvec) \: e^{- \delta U(\Rvec)},
\end{equation}
where $\Rvec=(R_0, R_1, ..., R_M=P R_0)$ is a particular world-line configuration with $R_j=\{\rvec_{1,j}, \rvec_{2,j}, ..., \rvec_{N,j}\}$ the coordinates of all $N$ particles in time slice $j$, and $\dR= dR_0 ... dR_{M-1}$. All permutations $P$ of the bosons are incorporated in the periodic boundary condition $R_M=P R_0$. We use the primitive approximation \cite{Ceperley95} where $U(\Rvec)$ contains only the inter-particle interaction
given by the Aziz potential \cite{Aziz79}. The kinetic term $T(\Rvec)$ is a product of free-particle propagators \cite{Ceperley95},
\begin{equation}
\label{eq:kin}
T(\Rvec) = \prod_{k=1}^{N} \prod_{j=0}^{M-1} (4 \pi \lambda^{(k)} \delta)^{-3/2} e^{ -\frac{(\rvec_{k,j+1} - \rvec_{k,j})^2}{4\lambda^{(k)} \delta} },
\end{equation}
where $\lambda^{(k)}= \hbar^2/(2 m^{(k)})$ depends on the mass of particle $k$ \footnote{The $^4$He and the $^3$He masses in units of Kelvin are $0.0825$K and $0.0622$K, respectively.}.

We cannot just add a substitutional $^3$He atom to the setup and wait for it to hop around, because the exchange amplitude between $^4$He and $^3$He atoms is very
small in solid $^4$He, $J_{34}\sim 10^{-4}$K \cite{Richards75} compared to the temperatures of interest ($T \sim 0.5$K).
A partial solution to this problem is to allow for a special Monte Carlo update, which relabels $^3$He
and $^4$He trajectories (a $^4$He trajectory which is not part of any exchange cycle is chosen at random) thus leaving the world line configuration unchanged. As both $^4$He and $^3$He interact via the same potential, only the kinetic part in Eq. \eqref{eq:Z} is affected by this update. Using Eq. \eqref{eq:kin}, the acceptance probability is
\begin{equation}
p_{ex}=\frac{T_\text{new}}{T}=
\min\{1,e^{-(l_3-l_4)(m_4-m_3)/\delta} \},
\end{equation}
with $l_i=\sum_{j=0}^{M-1} (\rvec_{k_i,j+1} - \rvec_{k_i,j})^2 \hbar^2/2$ and $m_i$ the mass, where the index $i=3,4$ refers to the $^3$He and $^4$He particle of the current update, respectively.
Typical acceptance ratios are of the order of $10^{-7}$
and thus prohibitively low.
The problem is that the $^3$He trajectory has a bigger fluctuation volume than the $^4$He one.

\begin{figure}
\centerline{\includegraphics[angle = 0, width=1\columnwidth]{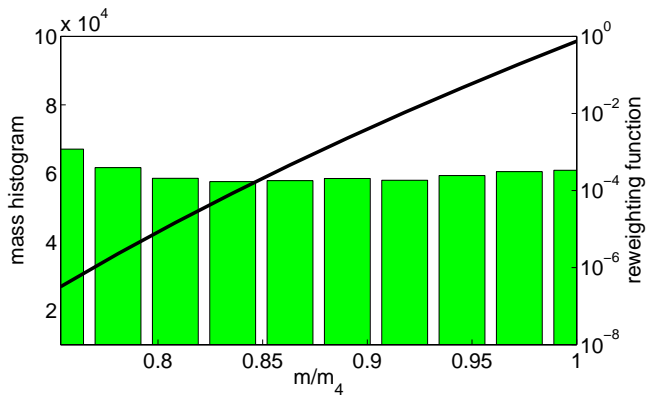}}
\caption{(Color online)
Reweighting function (line) and mass histogram (bars) as a function of ratio between the impurity particle mass and the $^4$He mass.
}
\label{fig:reweighting}
\end{figure}

To overcome this problem we introduce an update which gradually changes the mass $m$ of an impurity particle over the interval $[m_3,m_4]$. We do not allow for more than one impurity atom and work with a discrete set of 11 impurity masses $m=m_3+\Delta m *i$ where $i=0,1,\dots 10$ and $\Delta m=(m_4-m_3)/10$.
A gradual change in mass allows the crystal to relax and readjust the configuration
to the new impurity mass.
If $m$ is close or equal to $m_4$, the exchange updates are frequently accepted.
The quantities of interest are only measured in the "physical" sector, where $m=m_3$.
The acceptance probability for changing the mass from $m$ to $m \pm \Delta m$ is
\begin{equation}
p_{m \rightarrow m \pm \Delta m}=\min \{1,\left( 1\pm \Delta m/m \right)^{3M/2} e^{-l_3 \Delta m/\delta}\}.
\end{equation}

\begin{figure}
\centerline{\includegraphics[angle = 0, width=1\columnwidth]{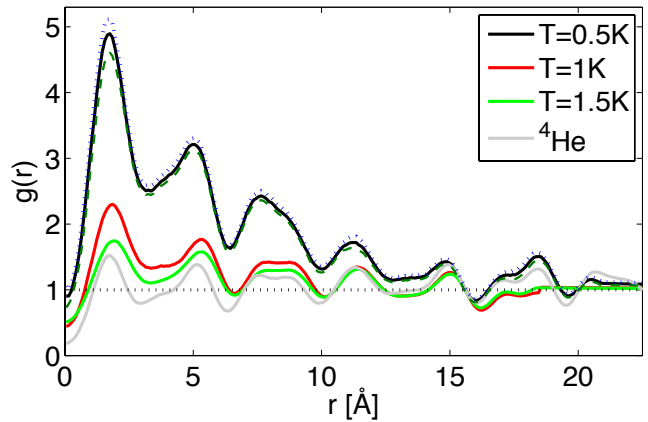}}
\caption{(Color online)
The probability density of finding a single $^3$He atom at a radial distance $r$ (cylindrical coordinates) from the core of the screw dislocation for different temperatures. Errors are indicated by the dashed lines for $T=0.5K$, and are of the same order at higher temperatures. The grey line shows the reference for $^4$He in the absence of an impurity. The density is $n = 0.0295$\Am3.
At large distances, $r > 30\text{\AA}$ ($r > 25\text{\AA}$ for $T\ge1$K), the $^4$He atoms are treated as inert particles (with fixed straight world lines) and form a zone inaccessible to the impurity atom.
}
\label{fig:gr}
\end{figure}

This in itself does not solve the problem, since on average
$p_{m \rightarrow m - \Delta m} < 1$ and we only rarely visit the low-mass sector of
the configuration space.
The final solution is in employing a reweighing (importance sampling) technique which ensures that
the probability of visiting different mass sectors are approximately equal. This is achieved by
introducing the reweighing function $g(m)$ shown in Fig. \ref{fig:reweighting} into the acceptance
probability
\begin{equation}
p_{m \rightarrow m \pm \Delta m} \rightarrow p_{m \rightarrow m \pm \Delta m} g(m) / g(m \pm \Delta m).
\end{equation}
This enables the impurity to efficiently sweep over the entire mass range.
Every time the impurity is a true $^3$He atom, we measure its distance $r$ to the nucleus
of the screw dislocations and update the histogram for the radial probability density $g(r)$
shown in Fig.~\ref{fig:gr}.

\begin{figure}
\centerline{\includegraphics[angle = 0, width=1\columnwidth]{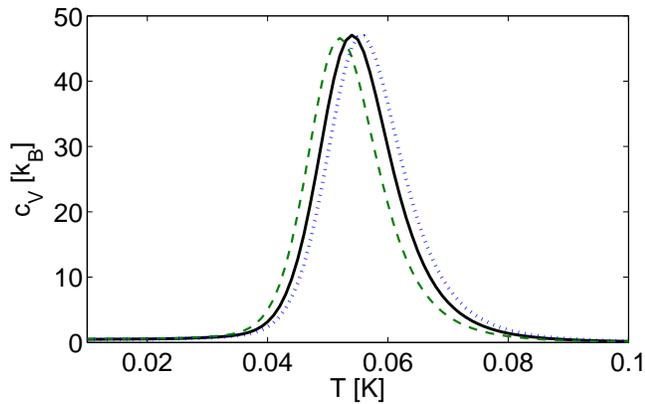}}
\caption{(Color online)
Specific heat as a function of temperature for a single $^3$He impurity in the strain field of a screw dislocation obtained directly from the $g(r)$ curve at $T = 0.5$K in Fig.~\ref{fig:gr} and assuming
$x_{\rm d} = 10^6 / a^2 = 7.6 \times 10^8$ cm$^{-2}$.
The specific heat is exponentially suppressed at temperatures well 
above where the  maximum is reached. The dashed lines correspond to the errors coming from the error on $g(r)$ in Fig.~\ref{fig:gr}.
The specific heat curves resulting from $g(r)$ at higher temperatures ($T=1$K and $T=1.5$K in Fig.~\ref{fig:gr}) are the same as the one shown, within error bars.
}
\label{fig:spec_heat}
\end{figure}

Relating $g(r)$ to the effective potential energy $E(r)$ between a $^3$He atom and the dislocation core is straightforward. The exchange matrix element between $^3$He and \he4 is negligible ($\sim 10^{-4}$K)
compared to the temperatures of interest ($T > 20$~mK) and can be ignored altogether,  leaving us with
the classical Boltzmann distribution $g(r)\propto \exp [ -\beta E(r) ]$. At large distances we
assume $E(r) \propto r^{-2}$ from elasticity theory~\cite{Andreev75} and proceed as follows:
first we we fit the tail of $g(r)$ to $g_\infty \exp ( -\beta B/r^2 )$ law to
determine the asymptotic behavior ($g_\infty$ and $B$ are fit parameters), and then
we obtain the potential energies from $E(r)=-T\ln [g(r)/g_\infty ]$  for distances $r < 20$\AA. The partition function is
found by integrating $\exp [ -\beta E(r) ]$ over all lattice sites up to some cut-off
value $r_{\rm max} = 1 / \sqrt{\pi x_{\rm d}} $ where $x_{\rm d}$ is the dislocation density per $a^2$
and $a$ the inter-particle distance. The specific heat $c_V$ shown in Fig.~\ref{fig:spec_heat},
is directly calculated from $g(r)$ shown in Fig.~\ref{fig:gr} with only $x_{\rm d}$
as a free parameter. The partition function can be written approximately as $Z\approx 1/x_{\rm d} + N_B \exp ( \beta E_B )$
where $N_B$ is the number of the deepest binding sites (with energy $-E_B \approx -T\ln [g_{max}/g_\infty ] = 0.8 \pm 0.1$K) per lattice period.

The specific heat maximum is roughly at $T_{\rm max} \approx E_B/ \ln [1/N_B x_{\rm d}]$) with only a logarithmic dependence on the free parameter $x_d$. The peak falls in the same temperature range as the peak of Ref.~\cite{Lin07}. The peak amplitude scales with the $^3$He concentration (since the Pauli exclusion can be neglected for low $x_3$ and assuming full equilibration)
and has the typical shape of a Schottky peak for a system with two degenerate energy levels (see above).
Further refinements of our model, such as working with discrete lattice points close to the core, using a Fermi function, or a distribution of binding energies for different dislocation types, are possible, but do not seem needed.

The Shevchenko state of the network of interconnected superfluid dislocations~\cite{Shevchenko87, Boninsegni07} predicts a crossover in the specific heat from $c_V \propto T$ above the bulk transition temperature $T_c$ to
$c_V \propto T^3$ at low temperatures $T < T_c$. This signal, however, might be extremely small and undetectable
leaving the $^3$He contribution as the leading one. Also, binding of $^3$He to dislocations
has an immediate effect on the shear modulus leading to crystal stiffening at $T<T_{\rm max}$ as observed in Ref.~\cite{Day07}.

\begin{figure}
\centerline{\includegraphics[angle = 0, width=1\columnwidth]{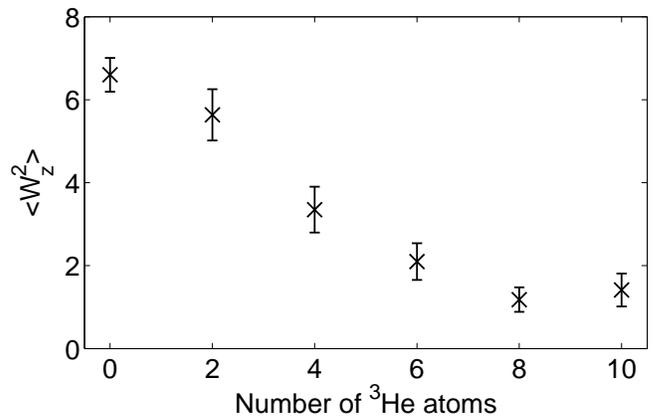}}
\caption {
The reduction in the winding numbers $\langle W_z^2 \rangle $ (proportional to the superfluid density) along the nucleus of the screw dislocation when the $^3$He concentration at the dislocation core
is increased. The temperature is $T = 0.5$K and  the density is $ n = 0.0295$\Am3.
}
\label{fig:bunch}
\end{figure}

Increasing the $^3$He concentration in the nucleus of the superfluid screw dislocation is expected to reduce the superfluid density. We assume that $^3$He atoms will cluster at the point where dislocations
intersect. We model their effect by introducing different numbers of impurities to the dislocation core next
to each other (in this simulation we do {\it not} employ any of the special updates and reweighing mentioned above and thus
the $^3$He cluster always remains in the core). We see in Fig.~\ref{fig:bunch} that about four $^3$He atoms
are required to suppress the superfluid response along the core. This mechanism may explain the reduction of the superfluid response for concentrations $x_3 > 1$ ppm observed in Ref.~\cite{Kim08}. 

Our last consideration is about the kinetic relaxation of $^3$He atoms. So far we assumed
thermodynamic equilibrium  which is not necessarily the case. NMR measurements established
that the tunneling motion of $^3$He atoms in $^4$He crystals is characterized by the hopping amplitude $J_{34} \sim 10^{-4}~$K.
Any strain field producing an energy level bias between the nearest neighbor sites $\xi \approx a(dE/dr)$
much larger than $z J_{34}$, where $z$ is the coordination number, will localize $^3$He atoms. To move around,
impurities have to exchange energy with the environment. At low temperatures, the leading mechanism
is provided by the one-phonon coupling and leads to hopping rates
$\tau ^{-1} \sim J_{34}^2 \xi^2 T/ \Theta_D^4$ where $\Theta_D$ is the Debye temperature \cite{KM}.
It is clear from the value of the binding energy that in the vicinity of the dislocation
core the condition $\xi \gg zJ_{34}$ is definitely satisfied. The slowest rate is for
$\xi \sim zJ_{34}$, \textit{i.e.} for $^3$He to cross the boundary between the band motion and localized states.
One can see, that the corresponding relaxation time is of the order of years at low temperature
leading to sample history dependent effects, as observed in experiments.

In conclusion, we have studied numerically the binding of $^3$He to the screw dislocation from first principles.
We find a binding energy of $0.8 \pm 0.1$K in agreement with published estimates. The binding of $^3$He impurity atoms to dislocation
cores at low temperature results in a specific heat peak in the same temperature interval as
observed experimentally in Ref.~\cite{Lin07, Lincorrection}, and may radically change superfluid properties of the
dislocation network even at minute $^3$He concentrations. Our data also provide quantitative support to the mechanism proposed in Ref.~\cite{Day07} as an explanation for the crystal stiffening at low temperature.

This work was supported by the National Science Foundation
under Grants Nos. PHY-0653183 and PHY-065135, and the Swiss National Science Foundation.
Simulations were performed on the Brutus cluster of ETHZ.
We acknowledge useful discussions and help from A. Kuklov. We thank Xi Lin and M. H. W Chan for showing us their most recent data.

\end{document}